\documentclass[superscriptaddress,aps,prl,twocolumn,showpacs]{revtex4}
\usepackage{graphicx}
\usepackage{amsmath,amssymb,bm}
\usepackage{color}

\begin{document}

\title{Testing black hole superradiance with pulsar companions}

\author{Jo\~ao G. Rosa \footnote{Also at Departamento de F\'{\i}sica e Astronomia, Faculdade de Ci\^encias da Universidade do Porto, Rua do Campo Alegre 687, 4169-007 Porto, Portugal.}}
\email{joao.rosa@ua.pt} 
\affiliation{Departamento de F\'{\i}sica da Universidade de Aveiro and CIDMA, Campus de Santiago, 3810-183 Aveiro, Portugal} 

\begin{abstract}

We show that the magnetic dipole and gravitational radiation emitted by a pulsar can undergo superradiant scattering off a spinning black hole companion. We find that the relative amount of superradiant modes in the radiation depends on the pulsar's angular position relative to the black hole's equatorial plane. In particular, when the pulsar and black hole spins are aligned, superradiant modes are dominant at large angles, leading to an amplification of the pulsar's luminosity, whereas for small angles the radiation is dominantly composed of non-superradiant modes and the signal is attenuated. This results in a characteristic orbital modulation of the pulsar's luminosity, up to the percent level within our approximations, which may potentially yield a signature of superradiant scattering in astrophysical black holes and hence an important test of general relativity.

\end{abstract}


\maketitle


It has been known since the 1970s that low-frequency waves can be amplified upon scattering off a rotating black hole (BH) \cite{Zeldovich, Starobinsky, Teukolsky:1973, Press:1973zz, Teukolsky:1974yv}. Superradiant scattering occurs for waves of frequency $\omega <m \Omega$, where $\Omega$ denotes the angular velocity of the BH horizon and $m$ the azimuthal number that characterizes the wave's angular momentum. Such waves carry negative energy and spin into the BH from the point of view of a distant observer, which is possible inside the ergoregion that surrounds a rotating BH. Such an observer thus sees the BH lose a fraction of its mass and angular momentum in the scattering process (see \cite{Brito:2015oca} for a recent review of this topic).

Superradiance is thus inherently associated to the space-time distortion produced by rotating BHs. Finding observational evidence for this process would then provide an extremely important test of Einstein's general relativity and a unique insight into the properties of the most compact objects in the universe.

Pulsar-BH binaries have long been regarded as the `holy grail' for testing the fundamental theory of gravity, combining one of the most accurate `clocks' in the universe with the strong space-time warping produced by a BH. Finding these systems is, in fact, one of the key goals of the Square Kilometer Array \cite{Aharonian:2013av}, as well as for current gravitational wave observatories such as Advanced LIGO/VIRGO \cite{Kalogera:2006uj}, where detection of up to a few binary mergers per year may be expected \cite{Singer:2014qca}. 

In this Letter, we show that the radiation emitted by a pulsar can undergo superradiant scattering off a rotating BH companion, constituting the first example of an astrophysical system where observational evidence for BH superradiance can be found.

We begin by considering the angular velocity of a Kerr BH of mass $M$ and spin $J=aMc$, which is given by:
\begin{eqnarray} \label{angular_velocity}
\Omega={ac\over {r_+^2+a^2}}\simeq 10\ \mathrm{kHz}\left({10M_\odot \over M}\right) \left({\tilde{a}\over 1+\sqrt{1-\tilde{a}^2}}\right)~,
\end{eqnarray}
where $r_+= (GM/c^2)(1+\sqrt{1-\tilde{a}^2})$ denotes the radial position of the outer horizon and $\tilde{a}= J c/ GM^2$ is the dimensionless spin parameter. This implies that, for stellar mass BHs in the range $10M_\odot -30 M_\odot$, superradiant amplification can only be significant for waves below the $1\ \mathrm{kHz}-10$ kHz frequency range, particularly since it is suppressed for large multipole numbers $l$ and $m$ \cite{Brito:2015oca}.

Such low frequency radiation is naturally produced by pulsars. These rapidly spinning neutron stars harbor extremely large magnetic fields formed in the collapse of the parent star. For ordinary pulsars, these can reach $10^8-10^9$ T, and even larger values for magnetars. This makes pulsars powerful emitters of magnetic dipole radiation at the rotation frequency $\omega_p\sim1\ \mathrm{Hz}-10\ \mathrm{kHz}$, the upper bound corresponding to the rapidly spinning millisecond pulsars (see e.g.~\cite{Kramer:2004gi}).

In addition, spinning neutron stars may also be an important source of gravitational waves due to deviations from axial symmetry. Several sources of asymmetry have been discussed in the literature, such as anisotropic stresses supported by the neutron star's solid crust, a misalignment between the rotation axis and the principal axis of inertia as a result of its violent formation, or a misalignment between the rotation and magnetic axes (see e.g.~\cite{Bonazzola:1995rb}). Due to its quadrupole nature, gravitational radiation is generically emitted at both the pulsar's angular frequency and twice its value, in amounts depending on the structure of the deformation.

It is thus clear that pulsars can emit both electromagnetic (EM) and gravitational waves (GWs) with frequency below the threshold for superradiant scattering, potentially providing two different channels for astrophysical observations. The question that remains to be addressed is whether this radiation has the correct multipolar character to extract energy and spin from a BH companion.


Let us first review the basic features of EM and GW scattering in the Kerr spacetime. A convenient framework to study these processes is the Newman-Penrose (NP) formalism \cite{Newman:1961qr}, where one projects the Maxwell and Weyl tensors describing the waves along a null tetrad. For the Kerr spacetime, a useful choice is the Kinnersley tetrad \cite{Kinnersley:1969zza}, where two of the 4-vectors coincide with the principal null directions of the Weyl tensor. This allows one to construct complex scalar functions, $\Upsilon_s$, known as the NP scalars and which encode the different spin components of the EM ($s=\pm 1$) and GW ($s=\pm 2$) radiation. 

These quantities satisfy independent scalar wave equations that can be solved using separation of variables \cite{Teukolsky:1973}. In particular, 
they admit a generic mode expansion in Boyer-Lindquist coordinates $(t,r,\theta,\phi)$ \cite{Boyer:1966qh} of the form:
\begin{eqnarray} \label{NP_mode_expansion}
\Upsilon_s&=&\sum_{l,m,\omega} e^{-i\omega t}e^{im\phi}{}_sS_{lm}(\theta){}_sR_{lm}(r)+ (\omega\rightarrow -\omega)
\end{eqnarray}
For simplicity, we will first focus on modes of the form $e^{-i\omega t}$ and discuss the $e^{i\omega t}$ modes below. The angular functions ${}_sS_{lm}(\theta)e^{im\phi}$ correspond to spin-weighted spheroidal harmonics, which reduce to the well-known spin-weighted spherical harmonics ${}_s Y_{lm}(\theta,\phi)$ \cite{Goldberg:1966uu} for $a\omega\ll 1$. The problem then reduces to a single radial wave equation, known as the Teukolsky equation \cite{Teukolsky:1973}:
\begin{eqnarray} \label{radial_equation}
\Delta{d^2{}_sR_{lm}\over dr^2}+2(s+1)(r-M){d{}_sR_{lm}\over dr}+\qquad\nonumber\\
\left({K^2-2is(r-M)K\over\Delta}+4is\omega r-{}_sA_{lm}\right) {}_sR_{lm}=0~,
\end{eqnarray}
where, in units such that $G=c=\hbar=1$, $K(r)=(r^2+a^2)\omega-ma$, $\Delta=r^2-2Mr+a^2$ and the angular eigenvalue can be written as a series expansion ${}_sA_{lm}=l(l+1)-s(s+1)+\sum_{k=1}^{+\infty}c_k(a\omega)^k$ \cite{Berti:2005gp}. 

Although no exact solution of this equation is known, approximate solutions can be found for low frequencies, $\omega \ll r_+^{-1}$, in two overlapping regions. These ``near" and ``far" solutions, valid for $r-r_+\ll \omega^{-1}$, and $r-r_+\gg r_+$, respectively, can then be matched to produce a complete solution. We then obtain a relation between the incoming and outgoing radiation, and consequently the energy flowing into the BH horizon. This is encoded in the gain/loss factor \cite{Starobinsky}:
\begin{eqnarray} \label{fractional_gain}
{}_sZ_{lm}&\equiv& {dE_{out}/dt \over dE_{in}/dt}-1= -{dE_{BH}/dt \over dE_{in}/dt}\nonumber\\
&\simeq& -\varpi{2\over \epsilon}(4\omega M \epsilon)^{2l+1}C_{ls}^2
\prod_{k=1}^l \left(k^2+{4\varpi^2\over \epsilon^2}\right)~,
\end{eqnarray}
where $\varpi=(\omega-m\Omega)r_+$, $\epsilon=(r_+-r_-)/2M$ and $C_{ls}=(l+s)!(l-s)!/ (2l)!(2l+1)!$. We thus conclude that superradiant scattering, ${}_sZ_{lm}>0$, occurs when $\omega<m\Omega$  for co-rotating wave modes, $m>0$. Also, amplification is larger for the lowest multipoles and, since $l\geq |s|$, it is more pronounced in the EM case (at low frequencies). 

We note that ${}_sZ_{lm}(\omega)={}_sZ_{l,-m}(-\omega)$ for arbitrary frequencies \cite{Teukolsky:1973}, so that superradiance will occur for $m<0$ for the wave modes of the form $e^{i\omega t}$.

Superradiant amplification is, however, much larger for higher frequencies, in particular for near-extremal BHs ($\tilde{a}\simeq 1$). In this regime, we need to employ numerical methods to solve the Teukolsky equation. A simple procedure, first used in \cite{Press:1973zz}, consists in numerically integrating an ingoing wave solution at the horizon up to a large distance, where one can extract the coefficients of the incoming and outgoing radiation, as we describe in detail in a companion article \cite{Rosa}. 

For GWs, we obtain a maximum amplification $Z^{max}\simeq 1.02$ for the $l=m=2$ mode with $\tilde{a}=0.999$ and $\omega\simeq 2\Omega$. In the EM case, the maximum gain is considerably lower, $Z^{max}\simeq 0.044$, for $l=m=1$, in agreement with \cite{Press:1973zz}. For non-superradiant modes, $Z<0$ and $|Z|$ increases with the frequency. In particular, the lowest non-superradiant multipoles approach the maximal absorption limit, $Z\simeq-1$, for $\omega<m\Omega$. In addition, for a given frequency, higher multipoles exhibit progressively smaller gain/loss factors, such that only a finite number of modes will effectively be relevant in a scattering problem (see also \cite{Dolan:2008kf}).

A physical wave, described by real fields, generically contains both $e^{\pm i\omega t}$ frequency modes, as well as different $(l,m)$ multipoles. Whether an overall amplification of the signal occurs upon scattering off a Kerr BH then depends on the relative amplitude of the different incoming modes, which we will now determine for the pulsar-BH binary.


We consider the limit where the orbital distance is large, $d\gg \lambda\gtrsim r_+, R_p$, where $\lambda$ is the wavelength of the radiation and $R_p$ the neutron star's radius. In this limit, it is a good approximation to consider the electromagnetic and gravitational fields generated by the rotating neutron star in flat space, which then give the incident wave for the scattering problem. 

Since in this limit the orbital period will largely exceed the pulsar's rotational period, we may first study the scattering problem for a given pulsar position $(d, \theta_p,\phi_p)$ in the BH frame and then include the orbital variation of these parameters.
 
It is also a sufficiently good approximation to treat the pulsar as a point-like spinning magnetic dipole and mass quadrupole. Here we will focus on binary systems where the pulsar and BH spins are either aligned or anti-aligned, although the same method can be applied to arbitrary configurations. 

For a misalignment angle $\alpha$ between the pulsar's spin and its magnetic dipole moment, we have:
\begin{eqnarray} \label{magnetic_moment}
\ddot{\mathbf{M}}_p={1\over2}M_0\omega_p^2\sin\alpha e^{-i\omega_p t}\left[\mathbf{e}_x\pm i\mathbf{e}_y\right]+c.c.~,
\end{eqnarray}
in the Cartesian coordinates associated with the Boyer-Lindquist frame, with the upper (lower) sign corresponding to an aligned (anti-aligned) binary system. We will also consider the case where any deformation rotates with the pulsar, such that its quadrupole moment corresponds to that of a deformed ellipsoid with principal axis of inertia at an angle $\beta$ from the spin axis. The resulting quadrupole tensor includes two frequency components, as anticipated above, yielding \cite{Bonazzola:1995rb}:
\begin{eqnarray} \label{quadrupole_tensor}
{\ddot{Q}_{ij}\over Q_0}\!&=&\!{c_\beta\over2} e^{-i\omega_p t}\!\!\left(
\begin{tabular}{c c c}
0 & 0 & $\pm i$\\
0 & 0 & -1\\
$\pm i$ & -1 & 0
\end{tabular}\right)\!+\!
s_\beta e^{-2i\omega_p t}\!\!\left(
\begin{tabular}{c c c}
1 & $\pm i$ &0\\
$\pm i$ & $-1$ & 0\\
0 & 0 & 0
\end{tabular}\right)\nonumber\\
&+&c.c.
\end{eqnarray} 
where $c_\beta\equiv\cos\beta$ and $s_\beta\equiv\sin\beta$. The amplitudes $M_0$ and $Q_0$ can be obtained from the properties of the neutron star but will not be relevant to our discussion. We note that the complex conjugate terms in Eqs.~(\ref{magnetic_moment}) and (\ref{quadrupole_tensor}) lead to waves of the form $e^{i\omega t}$.

From this we can use the standard dipole and quadrupole formulas to determine the EM and GW fields generated by the pulsar and the corresponding NP scalars. In the pulsar's frame, the $s<0$ NP scalars are, up to an $s-$dependent magnitude, given by:
\begin{eqnarray} \label{pulsar_NP}
\Upsilon_s\sim {e^{-i\omega (t-r)}\over r}{}_sY_{l_p,m_p}(\theta,\phi) + {e^{i\omega (t-r)}\over r}{}_sY_{l_p,-m_p}(\theta,\phi)\, ,
\end{eqnarray} 
where $l_p=1$ and $m_p=\pm 1$ for EM radiation of frequency $\omega_p$, while for GWs we have $l_p=2$ and $m_p=\pm 1, \pm2$ for $\omega = \omega_p, 2\omega_p$, respectively.

In the BH frame, however, the NP scalars become plane waves for large orbital distances, which are superpositions of all the different multipoles $l>|s|$. The multipolar decomposition of the NP scalars, which we will describe in detail in a companion article \cite{Rosa}, yields a remarkably simple result for the incoming radiation (directed towards the BH). For the $e^{-i\omega t}$ modes, we obtain for the incident power:
\begin{eqnarray} \label{incident_energy}
{dE_{in}^{(s,l,m)}\over dt}=\pi\left(\frac{\lambda}{d}\right)^2 \left|{}_{s}Y_{l_p,-m_p}(\theta_p)\right|^2\left|{}_{s}Y_{l,-m}(\theta_p)\right|^2P_s
\end{eqnarray}
where $P_s$ denotes the pulsar's total luminosity in EM or GWs, while for the $e^{i\omega t}$ modes we find an analogous result with $m_p\rightarrow -m_p$. 

\begin{figure}[htbp]
\centering\includegraphics[scale=0.9]{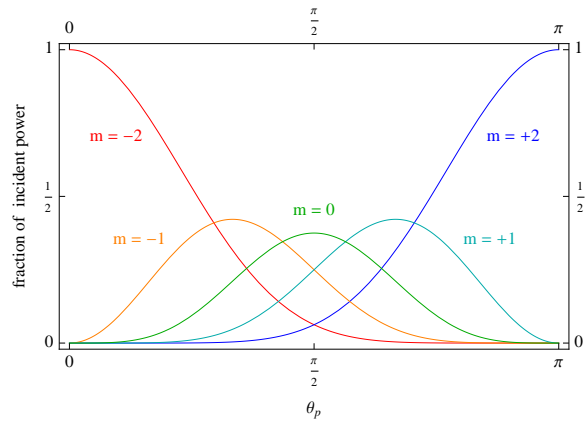}
\vspace{-0.5cm}
\caption{Relative fraction of the incident energy carried by the $l=2$ modes of the gravitational radiation emitted by the pulsar, normalized to the total quadrupole contribution, as a function of its polar angular position, $\theta_p$.}
\label{energy_fraction}
\end{figure}

We thus conclude that the fraction of the incident power carried by each $(l,m)$ mode depends on the polar angle $\theta_p$ but not on the azimuthal angle $\phi_p$, which only determines the phase of each mode. To better illustrate this, we show in Fig.~\ref{energy_fraction} the relative contribution of the different GW quadrupole modes.

As shown in this figure, the $m>0$ modes are dominant at large angles, while the $m<0$ modes give the largest contribution at small angles. Close to the BH's equator, $\theta_p=\pi/2$, most of the radiation is in the non-superradiant $m=0$ mode, and modes of opposite $m$ are present in equal amounts. This is a generic behavior, common to all multipoles in both the GW and EM cases.

Additionally, the relative contribution of the $e^{\pm i\omega t}$ modes also depends on the angle $\theta_p$ through distinct spin-$s$ harmonics, which correspond to the NP scalars in the pulsar's frame. Hence, the amount of (non-)superradiant modes in the incident radiation depends on the pulsar's angular position relative to the BH's rotation axis, and the radiation is generically ``polarized".

\begin{figure}[htbp]
\centering\includegraphics[scale=1]{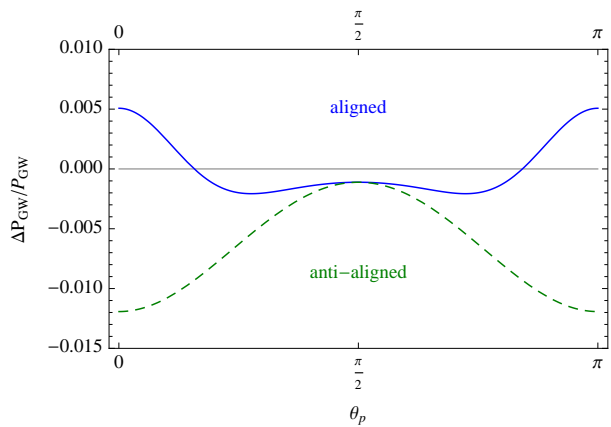}
\caption{Modulation of the pulsar's luminosity in GWs of frequency $\omega=2\omega_p\simeq 2\Omega$ with a near-extremal BH companion, $\tilde{a}=0.999$, as a function of the polar angle $\theta_p$. The solid blue (dashed green) curve corresponds to the case where the pulsar and BH spins are aligned (anti-aligned). The orbital distance is $d=10\lambda$ in this example.}
\label{angular_modulation}
\end{figure}

The fraction of the incident energy that is absorbed or amplified by the BH can be determined by multiplying the incident energy of each mode by the corresponding gain/loss factor ${}_sZ_{lm}$. We may then obtain the effective pulsar luminosity by subtracting the energy flowing into the BH per unit time from the total power. We illustrate this in Fig.~\ref{angular_modulation} for GWs of frequency $\omega=2\omega_p\simeq 2\Omega$ and $\tilde{a}=0.999$. This would correspond to e.g.~a millisecond pulsar around a 16$M_\odot$ near-extremal BH.

\vspace{0.5cm}

This figure shows that results depend on the relative orientation of the pulsar and BH spins. On the one hand, when the spins are aligned, superradiant modes of the form $e^{-i\omega t}$ and $m>0$ ( $e^{i\omega t}$ and $m<0$) are dominant at large (small) angles, and the signal is amplified when the pulsar's position makes an angle larger than $\sim 62^\circ$ with the equatorial plane, both above and below the equator. Closer to the equator, the incident radiation is dominated by non-superradiant modes and the pulsar luminosity decreases. We note that, in this example, only the $l=2, 3$ modes are relevant, with higher multipoles having negligible gain/loss factors.

On the other hand, when the spins are anti-aligned, the relative contributions of the  $e^{-i\omega t}$ and  $e^{i\omega t}$ modes are reversed. The incident wave is then dominated by non-superradiant modes and the radiation is mostly absorbed for all angles. In the equatorial plane, modes with opposite $m$ and $\omega$ are present in equal amounts and the signal is attenuated by the same factor in both the aligned and anti-aligned configurations. 

We find similar results for EM waves, although amplification is less pronounced ($\lesssim 10^{-4}$) and requires angular deviations from the equatorial plane of more than $77^\circ$. The example in Fig.~\ref{angular_modulation} corresponds to the most promising scenario, since amplification factors are maximal for GWs close to the superradiance threshold $\omega\sim 2\Omega$ and near-extremal BHs. For lower frequencies and BH spins we nevertheless find similar qualitative results, with e.g.~an overall modulation of $\sim 0.01$\% for $\tilde{a}=0.9$, for which we will give a detailed analysis in \cite{Rosa}.

\begin{figure}[t]
\centering\includegraphics[scale=0.99]{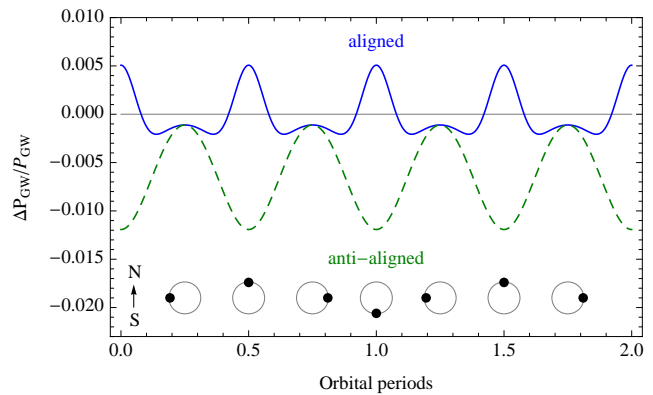}
\vspace{-0.3cm}
\caption{Modulation of a pulsar's GW luminosity for a polar circular orbit around a near-extremal BH ($\tilde{a}=0.999$), for frequency $\omega=2\omega_p\simeq 2\Omega$ and orbital radius $d=10\lambda$. The solid blue (dashed green) curve corresponds to the case where the pulsar and BH spins are aligned (anti-aligned). The bottom diagrams depict the pulsar's position in its orbit, with the arrow indicating the BH spin axis.}
\label{GW_modulation}
\end{figure}

Orbits in the Kerr space-time are generically non-planar, making the pulsar probe different polar angles in its orbit \cite{Singh:2014nta} and thus inducing a modulation of its luminosity. Here we consider the simplest case of planar circular orbits, which are a good approximation at large orbital radius. Since the effects of superradiance are more pronounced close to the poles, we illustrate in Fig.~\ref{GW_modulation} the best-case scenario of a polar orbit for the example considered above.

While we observe a single modulation of the luminosity for anti-aligned spins, an aligned binary system exhibits an additional modulation, which is entirely due to superradiant scattering when the pulsar moves away from the equatorial plane. This {\it double-modulation} is a generic feature, which is more or less pronounced depending on the radiation frequency, BH spin and orbital inclination, thus potentially yielding a `smoking gun' for superradiance.

Although we have considered the total pulsar luminosity, a similar modulation will occur along different lines-of-sight, as we will describe in detail in \cite{Rosa}. Note that incoming modes are further modulated by the corresponding spin-$s$ harmonics along a given direction, which may either enhance or reduce the signal modulation at different points in the orbit. The analysis given above nevertheless yields the generic magnitude and orbital evolution of the effects of superradiance. Weak lensing may also potentially pose an important experimental challenge along particular lines-of-sight \cite{Kruse:2014kaa}, although its effects may in principle be distinguished from superradiant amplification by measuring different EM and GW frequencies.


In the example above the effects of superradiance on the pulsar's luminosity are below the percent level. However, we expect this effect to become more pronounced for smaller orbital radii, namely for systems close to merging, where the incident power is larger. Moreover, the incident wave should approach the single multipole form obtained in the pulsar's frame as the radius decreases, thus favoring superradiant modes for an aligned binary. Although this lies outside the scope of this work, it motivates further study of this problem beyond the plane wave approximation. 

For EM waves, the modulation effect is quite suppressed and, furthermore, low-frequency radio waves are easily absorbed/attenuated in astrophysical plasmas. In fact, the plasma in the pulsar's magnetosphere may prevent the magnetic dipole radiation from reaching the BH. Even for pulsars where this plasma is periodically absent, such as B1931+24 \cite{Gurevich}, free-free absorption in the interstellar medium will most likely prevent the radiation from reaching the Solar System, and the superradiant modulation can at most be probed through its indirect effect e.g.~in heating intervening gas clouds \cite{Lacki:2010jr}.

GWs can yield a much stronger and cleaner evidence for superradiance in pulsar-BH binaries. The next generation of ground-based detectors, such as Advanced LIGO/VIRGO, should be sensitive to GWs emitted by deformed neutron stars, with current detectors already placing stringent constraints on pulsar ellipticities \cite{Aasi:2013sia}. Our results thus give additional motivation to search for pulsar gravitational radiation in such binary systems.

Evidence for BH superradiance would be a very important step in testing the yet unexplored corners of general relativity. Superradiance may also yield a unique laboratory for particle physics, through the potential instabilities that it may generate for ultra-light fields confined in the BH's gravitational potential \cite{Press:1972zz, Damour:1976, Zouros:1979iw, Detweiler:1980uk, Cardoso:2004nk, Furuhashi:2004jk, Cardoso:2005vk, Dolan:2007mj, Arvanitaki-JMR, Rosa:2009ei, Rosa:2011my, Pani:2012bp, Brito:2013wya, Brito:2015oca}. This can also lead to massive scalar `hair' around a Kerr BH \cite{Herdeiro:2014goa}.

The pulsar luminosity modulation found in this Letter provides the first of potentially other ways of testing the occurrence of superradiant scattering in astrophysical BHs and we hope that this motivates further exploration of this fascinating aspect of BH physics.


{\vskip 0.2cm}
\begin{acknowledgments}
{\bf Acknowledgments}

I would like to thank Carlos Herdeiro, Vitor Cardoso, Paolo Pani and Richard Brito for useful discussions and suggestions. This research was supported by the FCT grant SFRH/BPD/85969/2012 and partially by the grant PTDC/FIS/116625/2010, the CIDMA strategic project UID/MAT/04106/2013 and the Marie Curie action NRHEP-295189-FP7-PEOPLE-2011-IRSES.
\end{acknowledgments}


\end{document}